\documentclass[10pt,twocolumn,letterpaper]{article}

\usepackage{cvpr}              

\usepackage{graphicx}
\usepackage{amsmath}
\usepackage{amssymb}
\usepackage{booktabs}
\usepackage{comment}
\usepackage{multirow}
\makeatletter
\@namedef{ver@everyshi.sty}{}
\makeatother
\usepackage{tikz}
\usepackage{csquotes}
\usepackage{bm}
\usepackage{url}

\usepackage[pagebackref,breaklinks,colorlinks]{hyperref}

\usepackage[capitalize]{cleveref}
\crefname{section}{Sec.}{Secs.}
\Crefname{section}{Section}{Sections}
\Crefname{table}{Table}{Tables}
\crefname{table}{Tab.}{Tabs.}

\def\cvprPaperID{42} 
\def\confName{CVPR}
\def\confYear{2023}

\title{Towards Robust Image-in-Audio Deep Steganography}

\author{Jaume Ros$^{1,*}$\\
\and
Margarita Geleta$^{2,*}$\\
\and
Jordi Pons$^{3}$\\
\and
Xavier Giro-i-Nieto$^{1}$
\and
$^{1}$Universitat Polit\`ecnica de Catalunya    \qquad 
$^{2}$UC Berkeley\qquad 
$^{3}$Dolby Laboratories\\
}

\begin{document}
\twocolumn[{%
\renewcommand\twocolumn[1][]{#1}%
\maketitle

\begin{center}
    \centering
    \captionsetup{type=figure}
    \includegraphics[width=\textwidth]{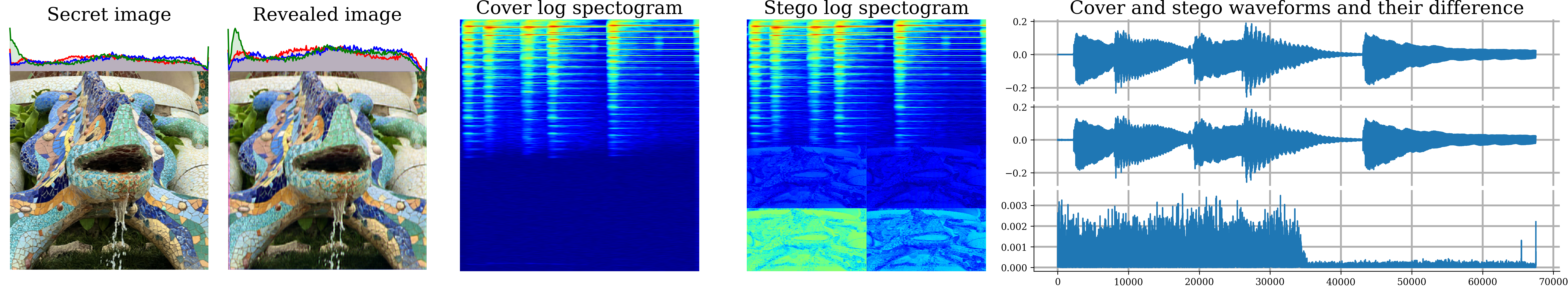}
    \captionof{figure}{\textbf{Components of the steganographic pipeline}. At the left, the \emph{secret} image to be embedded and the revealed image at the decoder end are shown together with their RGB density histograms, followed by the \emph{cover} log spectrogram and the \emph{stego} (\emph{cover} + embedded \emph{secret}) log spectrogram. At the right, the \emph{cover} and \emph{stego} waveforms are shown after applying inverse STFT, and their $L_1$ distance is computed. }
    \label{fig:teaser}
\end{center}%
}]

\def\thefootnote{*}\footnotetext{Indicates equal contribution.}\def\thefootnote{\arabic{footnote}}
\def\thefootnote{**}\footnotetext{The code is available for download on \url{https://github.com/migamic/PixInWav2}.}

\def\checkmark{\tikz\fill[scale=0.4](0,.35) -- (.25,0) -- (1,.7) -- (.25,.15) -- cycle;}

\begin{abstract}
The field of steganography has experienced a surge of interest due to the recent advancements in AI-powered techniques, particularly in the context of multimodal setups that enable the concealment of signals within signals of a different nature. The primary objectives of all steganographic methods are to achieve perceptual transparency, robustness, and large embedding capacity --- which often present conflicting goals that classical methods have struggled to reconcile. This paper extends and enhances an existing image-in-audio deep steganography method by focusing on improving its robustness. The proposed enhancements include modifications to the loss function, utilization of the Short-Time Fourier Transform (STFT), introduction of redundancy in the encoding process for error correction, and buffering of additional information in the pixel subconvolution operation. The results demonstrate that our approach outperforms the existing method in terms of robustness and perceptual transparency$^{**}$.
\end{abstract}

\section{Introduction}
The use of deep learning in steganography is a relatively new area of research, but it has already improved the performance of traditional steganographic techniques and opened up new possibilities for covert communication \cite{im_in_im_baluja, Baluja2020HidingIW, cnn_steganogrpahy, steg_net, im_in_im_skip, unet_image_steganography}. In particular, steganography has evolved to multimodal setups, enabling the embedding of signals of one modality within signals of a different modality \cite{hidden, hide_and_seek, geleta2021pixinwav}. 
In this paper, we present a novel approach to improving an existing image-in-audio deep steganography method \cite{geleta2021pixinwav}. Our approach involves a series of key enhancements, including the use of short-time Fourier transform (STFT) instead of the short-time discrete cosine transform (STDCT), introducing redundancy in the encoding process for error correction, and buffering additional information in the pixel subconvolution operation, among others. Our enhanced method effectively equips the steganographic agent with a powerful new set of tools to operate with. 

Our training pipeline hides ImageNet images \cite{deng2009imagenet} within audio samples from the FSDNoisy18K dataset \cite{fsdnoisy18k} (Figure \ref{fig:teaser}). To evaluate the effectiveness of our approach, we conducted comprehensive experiments and comparison with the baseline model using perceptual metrics for image quality (SSIM and PSNR) and audio quality (SNR). The results demonstrate that our approach outperforms the existing method in terms of robustness and perceptual transparency. 

To summarize, we make the following contributions:
\begin{enumerate}
    \item We improve the performance of the steganographic method in \cite{geleta2021pixinwav} by changing the real short-time discrete cosine transform (STDCT) by the complex short-time Fourier transform (STFT) and we show that increasing the resolution of spectral representation of the audios improves the model performance.
    \item We improve the \emph{secret} image reconstruction by introducing new image-in-audio replication-based embedding methods. Additionally, introducing redundancy via replication serves as error correction, enhancing the robustness of the method.
    \item We enhance the architectural design of the model by buffering the luma component of the YCbCr color representation of the \emph{secret} image in the subpixel convolution operation.
\end{enumerate}

\nocite{geleta2021pixinwav}

\vspace{-1mm}
\section{Related Work}

Steganography is the practice of concealing a \emph{secret} signal (which may be covert communication or a watermark) within a \emph{cover} (or \emph{host}) signal (the medium containing both signals is called \emph{stego} signal) with the objective of: (1) maximizing the perceptual transparency, i.e. maximizing the similarity between the \emph{cover} and \emph{stego} signals; (2) maximizing the robustness, which is the ability to withstand intentional or accidental attacks, and (3) maximizing the embedding capacity, the secret message size per 1-time or -space unit \cite{bender_steganography, exploring_steganography, hide_and_seek, information_hiding_survey}. The earliest known use of steganography dates back to ancient Greece \cite{conway2003code, siper2005rise}. In the modern era, steganography has been used for a variety of purposes, including copyright protection/watermarking \cite{bender_steganography, info_hiding_stego, cox1997secure, brassil1995hiding, hartung1999spread, mpeg2_watermarking, kundur1998digital, nikolaidis1996copyright}, military communications \cite{thicknesse1772treatise, siper2005rise, information_hiding_survey, Kahn1968TheC}, and feature tagging \cite{lin1999review, bender_steganography, caronni1995assuring, electronic_marking, tanaka1990embedding}. Recently, steganography found its connectionist approach and gave rise to a new area in deep learning known as \emph{deep steganography}.

First attempts in \emph{deep steganography} concerned unimodal setups, in which both the \emph{secret} and \emph{cover} signals are of the same modality, such as image-in-image or audio-in-audio. We find numerous examples of image-in-image steganography: 
Baluja \cite{im_in_im_baluja, Baluja2020HidingIW} incorporated a convolutional neural network with inception-like modules to encode a \emph{secret} full-sized color image in a dispersed a manner throughout the bits of the \emph{cover} image of the same size. Inspired by auto-encoding networks for image compression, the system learns to compress and place the \emph{secret} image into the least noticeable portions of the \emph{cover} image. For that aim, firstly, the network extracts features from the \emph{secret} image and merges them with the \emph{cover} image in the hiding step. Besides, the author mentions that this technique could be applied on audio samples by interpreting their spectrograms as images; Rehman \emph{el al.} \cite{cnn_steganogrpahy} employ a two-branch encoder to gradually extract features from both \emph{secret} and \emph{cover} images and syncronize them at several steps to produce \emph{stego} images; StegNet \cite{steg_net, im_in_im_skip} continues in the encoder-decoder line but improves the perceptual transparency of the method by introducing skip connections \cite{he2016deep} and separate convolutions in the architecture to improve convergence \cite{chollet2017xception}; Duan \emph{et al. } \cite{unet_image_steganography} propose concatenating both \emph{secret} and \emph{cover} signals and use U-Nets as hiding and revealing networks. At less extent, we find audio-in-audio steganography \cite{au_in_au}. 

More recently, researchers have also explored the use of deep learning for multimodal steganography, in which the \emph{secret} message and the \emph{cover} media are of different modalities. This approach has the potential to further improve the transparency and robustness of \emph{secret} messages, as well as expanding the range of \emph{cover} media that can be used for steganography. HiDDeN \cite{hidden} uses a convolutional decoder-encoder architecture to embed a secret string message within an image, unconstraining the \emph{secret} to be a specific kind of signal since a string of bits could represent any type of data. 
There have been attempts to address the specific case of image-in-audio steganography with classical methods \cite{im_in_au_2, im_in_au_3, im_in_au_4}, but leveraging the audio representation for embedding image data  has not been explored extensively in the deep learning context. One of such deep steganographic models is PixInWav \cite{geleta2021pixinwav}.  






\def\thefootnote{*}

\section{Preliminaries}
In this section we detail specific parts of interest of the PixInWav \cite{geleta2021pixinwav} model, that serves as a baseline for this work. 
PixInWav is a deep steganographic model for image-in-audio concealment. 
The pipeline is trained end-to-end and the trainable part consists of two U-Net-style networks: a \emph{hiding} network and a \emph{revealing} network.


The pipeline takes a \emph{secret} image $s$ and a \emph{cover} audio waveform $w$. First, $w$ is transformed into a spectrogram $M$ using the short-time discrete cosine transform (STDCT). Then, the \emph{hiding} network is applied on the pixel-shuffled $s$ for conversion into a low-power spectrogram watermark, which is residually added onto $M$, resulting in the \emph{stego} spectrogram $M'$. $M'$ can be transformed back to the temporal domain, $w'$, via inverse STDCT for transmission. At the decoder end, the \emph{revealing} network takes as input $M'$ to extract the revealed image $s'$ by pixel-unshuffling the network output. 

The loss function is a convex combination involving image and spectrogram reconstructions, with the addition of soft dynamic time warping (DTW) discrepancy \cite{softdtw} (with a smoothing parameter $\gamma$) between the \emph{cover} waveform and the \emph{stego} waveform for better temporal alignment.

\begin{equation}
\begin{aligned}
    \mathcal{L}(s,s', w,w', M,M') &= \beta\|s-s'\|_{1}  + \lambda\text{dtw}(w,w', \gamma=1) \\&+(1-\beta)\|M-M'\|_{2} 
\end{aligned}
\end{equation}

We refer the interested reader to the PixInWav paper \cite{geleta2021pixinwav} paper for details.

\textbf{Limitations}. Even though PixInWav has shown the feasibility of image-in-audio connectionist steganography, it suffers from the lack of robustness. 
This weakness stems from several architectural decisions. 

The first one is the uselessness of zero-padding in the pixel-shuffle operation. As in PixInWav \cite{geleta2021pixinwav}, pixel-shuffle (or sub-pixel convolution \cite{pixelshuffle}) is used to flatten the RGB image into a single channel by arranging the 3 color channels of each pixel side by side in a $2\times2$ grid, padding an empty element of value 0 as the fourth element. We show that this fourth element can serve as a buffer to transmit useful information of the image to encode allowing for better image reconstruction. 

The second weakness is the forgoing of image replication in the \emph{stego} spectrogram. The shape of the \emph{cover} spectrogram does not need to match the shape of the image. While images in the dataset are of size $256\times256$ ($512\times512$ after applying the pixel-shuffle operation), the spectrograms are 
larger in size and, in general, present a non-square shape\footnote{Geleta et al. \cite{geleta2021pixinwav} used spectrograms of shape $4096\times1024$. As mentioned in Section \ref{section:large-container}, these values can be arbitrarily increased or decreased. We chose to use a shape of $1024\times512$ for easier comparison with our STFT implementation and reduced computational load.}. This mismatch is overcome by stretching the image via bilinear interpolation, which is a reversible operation that allows to easily resize the image to any desired shape. However, the extra space can be exploited to replicate the \emph{secret} image, a procedure that can be deemed as an error correction technique for robustness improvement. 

Finally, the basis function of the short-time transform is type-2 discrete cosine transform (DCT) and the resolution is fixed. The choice of the STDCT over short-time Fourier transform (STFT) was reasoned in \cite{geleta2021pixinwav} because it had a smaller set of components: STDCT, being a real transform, results in a single real-valued spectrogram, in contrast to the complex STFT transform, that is decomposed in both, magnitude and phase. However, increasing the set of components increases the number of strategies to embed our \emph{secret} image. Additionally, the resolution of the transform can be increased arbitrarily up to computational constraints.

\section{Enhancements} 
We propose several enhancing features for the image-in-audio steganographic method proposed by PixInWav \cite{geleta2021pixinwav}. 

\begin{figure*}[ht!]
\centering
\includegraphics[width=\linewidth]{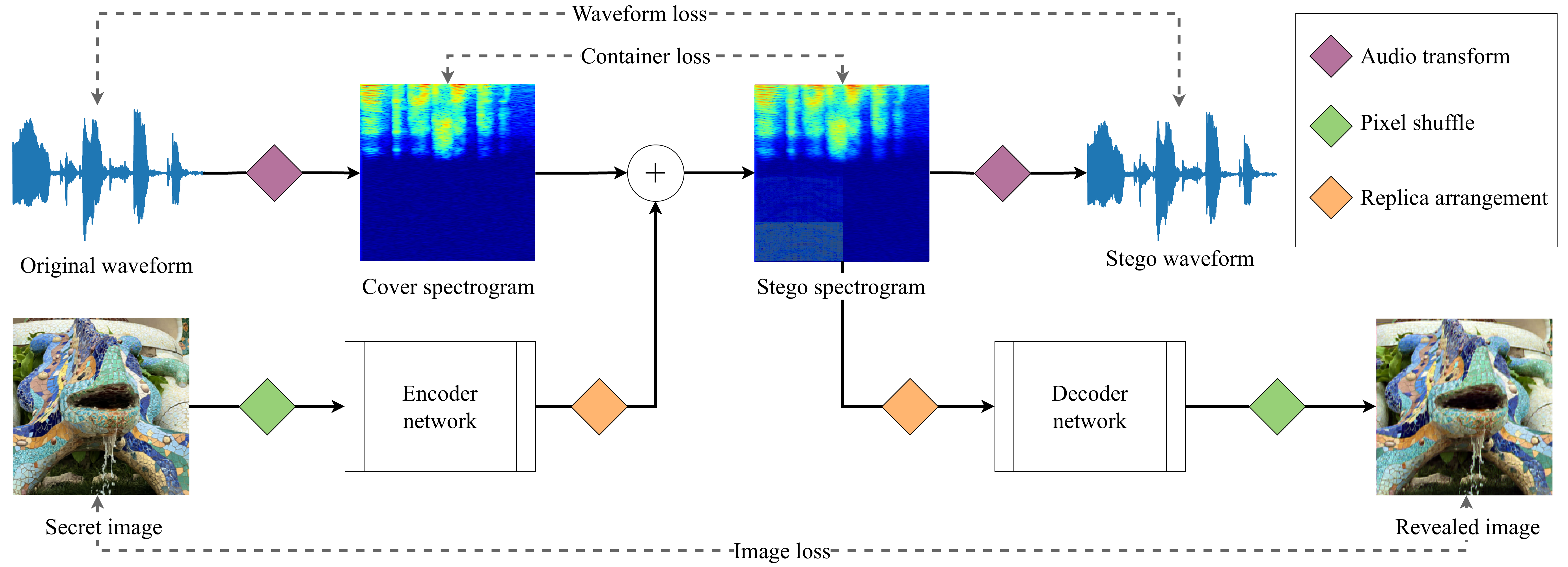} \caption{\textbf{Architecture of the proposed model}. The legend shows the three basic components of the steganographic pipeline that have been improved in our work: (pink) change of the audio transform, (green) buffering of luma in the pixel shuffle operation, and (orange) addition of different embedding methods based on replication.}
\label{fig:arch}
\end{figure*} 

\begin{figure*}[ht!]
\centering
\includegraphics[width=\linewidth]{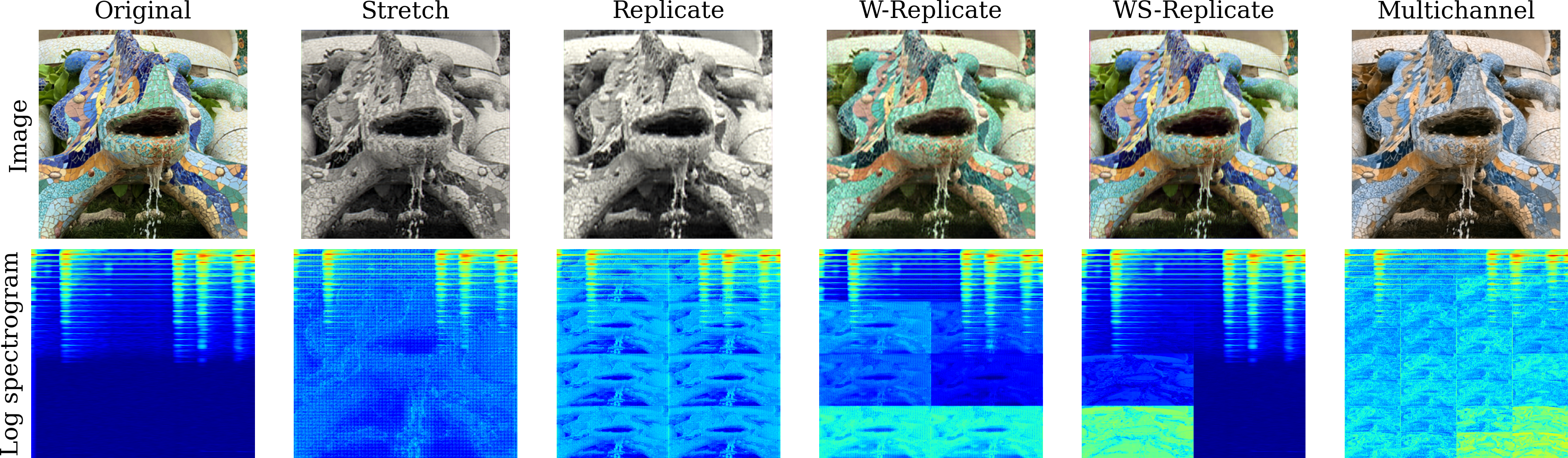}
\caption{\textbf{Comparison of different embedding methods operating in the STFT magnitude}. The first row represents the reconstructed image, while in the second row we can visualize the log spectrogram of the $2048 \times 1024$ \emph{stego} signal. Each of these models used $\beta=0.75$.}
\label{fig:embedding-methods}
\end{figure*}

\subsection{STFT instead of STDCT}
\label{sec:phase-loss-wav}

Given a time domain sequence $x[n]$, its discrete short-time transform operation is given by Equation \ref{eq:short-time}:
\begin{equation}
X_{\mathcal{T}}\{x[n]\}(m,F_k)=\sum_{n=mr}^{mr+N-1}x[n]\cdot h[n-mr] \cdot \mathcal{T}[n,F_k]
\label{eq:short-time}
\end{equation}
which is a bivariate function representing the energy of the $F_k$-th frequency component (equispaced frequency samples $F_k = k/N$) in the $m$-th frame. The time index is represented by $n$; $h$ is the low pass window function, and $r$ represents the hop size of the short-time transform. The function $\mathcal{T}[n,F_k]$ is the basis function of the transformation.

The STDCT, used by PixInWav, is a real transform using the type-2 DCT basis function as $\mathcal{T}[n,F_k]$, which produces a single 2D spectrogram from a 1D audio waveform. On the other hand, the STFT is a complex transform using the complex exponential function as its basis function  $\mathcal{T}[n,F_k] = e^{-i2\pi F_k n}$, which results in a complex signal that can be split into two 2D signals: the magnitude and the phase. We propose using the STFT instead of STDCT, where this duality of the \emph{cover} signal allows for more possibilities in how the \emph{secret} signal can be embedded onto the \emph{cover} signal. Both magnitude and phase, can be used as \emph{stego} signals in the same manner that the single spectrogram from the STDCT has been used. In this work we consider using any of the two signals as a single \emph{stego} signal, or using both of them at the same time.

\textbf{STFT magnitude as a single \emph{stego} signal}. One can approach embedding the information in the STFT magnitude in the same way as using a single STDCT spectrogram, with the main difference being that the imaginary part of the \emph{cover} signal remains unmodified. Since we are not directly distorting the phase component, the alignment performed by DTW can be deemed redundant. We propose changing the soft DTW used in \cite{geleta2021pixinwav} by a simpler $L_1$ distance between the \emph{cover} and \emph{stego} waveforms (Equation \ref{eq:loss0}):
\begin{equation}
\begin{aligned}
    \mathcal{L}(s,s', w,w',M,M') &= \beta\|s-s'\|_{1} + \lambda\|w-w'\|_{1} \\&+(1-\beta)\|M-M'\|_{2}
\end{aligned}
\label{eq:loss0}
\end{equation}

\textbf{STFT phase as a single \emph{stego} signal}. Since the phase has the same spatial dimensions as the magnitude signal, the same methods can be applied only on the phase component. In this case, only the imaginary part is modified.

\textbf{STFT magnitude and phase as \emph{stego} signals}. A more advanced setup has been developed in which both the STFT magnitude and phase can jointly serve as a \emph{stego} signal. To handle multiple \emph{stego} components, the architecture requires an adaptation: the two \emph{stego} components should be treated separately due to their very different structure. The proposed architecture uses different encoders and decoders for each \emph{stego} component. The two revealed images are fed into a third network that processes them to obtain a single image as output. Out of the multiple solutions tried, a simple trained weighted average worked the best. The loss function has been accommodated for the possibility of using multiple containers (Equation \ref{eq:new-loss}):
\begin{equation}
\begin{aligned}
    \mathcal{L}(&s,s',w, w', M,M', P, P') = \beta\|s-s'\|_{1} + \lambda\|w-w'\|_{1}\\&+(1-\beta) \left[(1 - \theta) \|M-M'\|_{2} + \theta \|P-P'\|_{2} \right] 
\end{aligned}
\label{eq:new-loss}
\end{equation}
where $M$ and $P$ now denote the magnitude and phase signals, respectively, and $M'$ and $P'$ correspond to their respective \emph{stego} components. These components are weighted by a new hyperparameter $\theta$, that controls the trade-off between magnitude and phase distortion. Notice that the waveform $w$ is still unique.

\subsection{Spectrogram replicas}\label{sec:embeddings}

We consider the case of encoding a $256 \times 256 \times 3$ RGB image (flattened to $512 \times 512$ after applying the pixel-shuffle operation) onto a $1024 \times 512$ spectrogram. 

PixInWav \cite{geleta2021pixinwav} made use of bilinear interpolation for upsampling the encoded image to match the spectrogram shape, only to be later downsampled to its original size before decoding. This strategy, \textit{Stretch} from now on, makes the encoding and decoding processes independent of the \emph{stego} size. However, other options could be devised. In this section we propose alternative architectures to address this problem (Figure \ref{fig:embedding-methods}), that make a better use of the space available by encoding multiple copies of the encoded image, and improving the \emph{secret} reconstruction and increasing the robustness of the steganographic method.


\textbf{\emph{Replicate} method}. Our simplest approach uses the fact that the \emph{cover} spectrogram is 
significantly larger than the \emph{secret} image and this allows for a natural replication of the encoded image that is added onto the host signal. When decoding, the two copies are jointly forwarded through the network, split and averaged to produce the final \emph{revealed} image.


\textbf{\emph{Weighted Replicate} method}. \textit{Weighted Replicate} (\emph{W-Replicate}) improves the previous method by scaling each replica by a trainable weight before adding them onto the container spectrogram, and also when merging them into a single one (essentially, a trained weighted average); resulting in a total of four trainable weights that are added to the model. This change allows the model to learn in which half of the STFT spectrogram (high or low frequencies) the information can be added causing the least distortion.


\textbf{\emph{Weighted} \& \emph{Split Replicate} method}.
The previous two methods decode the spectrogram directly, i.e. a tensor of shape $1024 \times 512$, with the two replicas side by side; this forces the network to treat both replicas equally. \textit{Weighted \& Split Replicate} (\emph{WS-Replicate}), improves upon this issue by first splitting the container signal and decoding the two replicas separately (i.e. concatenated in a 3rd dimension, resulting in a tensor of shape $512 \times 512 \times 2$). The encoder structure is the same as in \emph{W-Replicate}.


\textbf{\emph{Multichannel} method}. All the previous methods rely on the pixel-shuffle operation to flatten the image into a single color channel. \textit{Multichannel}, however, omits this step and has the model learn to encode the three color channels in the different replicas. Thus, the encoded image is of shape $256 \times 256 \times C$, being $C$ the desired number of output channels. Since eight $256 \times 256$ replicas can fit into a $1024 \times 512$ host signal, we set $C=8$. They are arranged in a $4 \times 2$ grid. As with \emph{WS-Replicate}, the decoder is fed on the replicas already split and concatenated, only that this time the output is already the final $256 \times 256 \times 3$ RGB image.
\begin{table*}[]
\resizebox{\linewidth}{!}{\begin{tabular}
{p{0.3cm}|p{7.25cm}|p{2.1cm}|p{0.5cm}|p{0.8cm}|p{1.75cm}|p{1.25cm}p{1.25cm}p{1.75cm}|p{1.25cm}p{1.5cm}}
\specialrule{.2em}{.1em}{.1em}
 \# & Model & Embedding method & $\beta$ & Luma & Container size & Revealed SSIM $\uparrow$& Revealed PSNR $\uparrow$ &Color restoration& Stego SNR $\uparrow$ & Waveform loss $\downarrow$ \\
\specialrule{.2em}{.1em}{.1em}
1 & Baseline PixInWav  \cite{geleta2021pixinwav} (STDCT, DTW $\lambda=10^{-4}$)& Stretch &   0.05   &   --       & $1024 \times 512$ & 0.84 & 23.80 & Full color &-3.08  &    10.53  \\
\hline
2 & PixInWav \cite{geleta2021pixinwav} (STDCT, DTW $\lambda=10^{-4}$) & Stretch &   0.01   &   --       & $1024 \times 512$ & 0.76 & 20.45 & Partial color & 8.72  &    1.05\\

3 & PixInWav \cite{geleta2021pixinwav} (STDCT, DTW $\lambda=10^{-4}$) & Stretch &   0.5   &   --       & $1024 \times 512$ & \textbf{0.86} & 25.29 & Full color & -14.90  &    89.11   \\

4 & PixInWav \cite{geleta2021pixinwav} (STDCT, DTW $\lambda=1$) & Stretch &   0.05   &   --       & $1024 \times 512$ & 0.39 & 11.17 & No color & 45.98  &    $1.1 \times 10^{-4}$  \\

5 & Modified PixInWav (STDCT, $L_1 \lambda=1$) & Stretch &   0.05   &   --       & $1024 \times 512$ & 0.37 & 10.84 & None & 34.65  &    $2 \times 10^{-5}$  \\
\hline
6 & Ours (STFT: magnitude, DTW $\lambda=10^{-4}$) & Stretch &   0.75   &   --       & $1024 \times 512$ & 0.73 & 20.06 & No color & 41.17  &    $4.8 \times 10^{-3}$   \\

7 & Ours (STFT: magnitude, $L_1 \lambda=1$) & Stretch &   0.85   &   --       & $1024 \times 512$ & 0.73 & 20.09 & No color & 43.27  &    $1.3 \times 10^{-4}$   \\

8 & Ours (STFT: magnitude, $L_1 \lambda=1/2$) & Stretch &   0.75   &   --       & $1024 \times 512$ & 0.69 & 20.58 & No color & 44.72  &    $1.2 \times 10^{-4}$   \\

9 & Ours (STFT: magnitude, $L_1 \lambda=1$) & Stretch &   0.75   &   --       & $1024 \times 512$ & 0.64 & 20.95 & Partial color & 44.66  &    $1.2 \times 10^{-4}$  \\

10 & Ours (STFT: phase, $L_1 \lambda=1$) & Stretch &   0.75   &   --       & $1024 \times 512$ & 0.52 & 14.89 & No color & 21.84  &    $2.1 \times 10^{-3}$   \\

11 & Ours (STFT: magnitude + phase, $L_1 \lambda=1$) & Stretch &   0.75   &   --       & $1024 \times 512$ & \textbf{0.87} & 26.27 & Partial color & 22.19  &    $7.2 \times 10^{-4}$  \\

12 & Ours (STFT: magnitude, $L_1 \lambda=1$) & Replicate &   0.75   &   --       & $1024 \times 512$ & 0.71 & 22.84 & Full color & 42.76  &    $1.6 \times 10^{-4}$\\

13 & Ours (STFT: magnitude, $L_1 \lambda=1$) & W-Replicate &   0.75   &   --       & $1024 \times 512$ & 0.64 & 20.00 & Full color & 38.25  &    $2.6 \times 10^{-4}$   \\

14 & Ours (STFT: magnitude, $L_1 \lambda=1$) & WS-Replicate &   0.75   &   --       & $1024 \times 512$ & 0.81 & 25.33 & Full color & 40.60  &    $1.9 \times 10^{-4}$   \\

15 & Ours (STFT: magnitude, $L_1 \lambda=1$) & Multichannel &   0.75   &   --       & $1024 \times 512$ & \textbf{0.87} & 24.08 & Partial color & 15.83  &    $3.3 \times 10^{-3}$  \\

\hline
16 & Ours (STFT: magnitude, $L_1 \lambda=1$) & Stretch &   0.75   &   --       & $2048 \times 1024$ & 0.70 & 19.92 & No color & 51.43  &    $5.6 \times 10^{-5}$   \\

17 & Ours (STFT: magnitude, $L_1 \lambda=1$) & Stretch &   0.75   &   \checkmark      & $2048 \times 1024$ & 0.71 & 19.99 & No color & 50.94  &    $6.2 \times 10^{-5}$  \\

18 & Ours (STFT: magnitude, DTW $\lambda=10^{-4}$) & Stretch &   0.75   &   \checkmark      & $2048 \times 1024$ &  0.79 &  20.63&  No color & \textbf{52.84} & $3.7 \times 10^{-4}$\\

19 & Ours (STFT: magnitude + phase, $L_1 \lambda=1$) & Stretch &   0.75   &   \checkmark      & $2048 \times 1024$ &  \textbf{0.91} &  28.35 & Full color  &  8.14 & $1.3 \times 10^{-3}$\\

\hline
20 & Ours (STFT: magnitude, $L_1 \lambda=1$) & Replicate &   0.75   &   --       & $2048 \times 1024$ & 0.68 & 22.30 & Partial color & \textbf{53.14}  &    $4.6 \times 10^{-5}$\\

21 & Ours (STFT: magnitude, $L_1 \lambda=1$) & Replicate &   0.75   &   \checkmark  & $2048 \times 1024$ & 0.73 & 20.27 & No color & 49.49 &    $6.9 \times 10^{-5}$ \\

22 & Ours (STFT: magnitude, DTW $\lambda=10^{-4}$) & Replicate &   0.75   &   \checkmark      & $2048 \times 1024$ &  0.64 &  19.67&  No color&   \textbf{51.94}& $4.3 \times 10^{-4}$ \\

\hline
23 & Ours (STFT: magnitude, $L_1 \lambda=1$) & W-Replicate &   0.75   &   --       & $2048 \times 1024$ & 0.83 & 26.09 & Full color & \textbf{55.22}  &    $4.1 \times 10^{-5}$  \\

24 & \textbf{Ours (STFT: magnitude, $L_1 \lambda=1$)} & \textbf{W-Replicate} &   0.75   &  \checkmark   & $2048 \times 1024$ & \textbf{0.88} & 23.81 & Full color & \textbf{50.46}  &    $6.8 \times 10^{-5}$  \\

25 & Ours (STFT: magnitude, DTW $\lambda=10^{-4}$) & W-Replicate &   0.75   &   \checkmark      & $2048 \times 1024$ &  0.79&   21.05 & No color & 37.62  & $1.2 \times 10^{-2}$ \\

26 & Ours (STFT: magnitude, $L_1 \lambda=1$) & W-Replicate &   0.5   &   \checkmark      & $2048 \times 1024$ &  0.77&  20.65& No color& \textbf{54.12}& $3.8 \times 10^{-5}$\\

27 & Ours (STFT: magnitude, $L_1 \lambda=1$) & W-Replicate &   0.9   &   \checkmark      & $2048 \times 1024$ & 0.82 &  25.24& Full color &  43.31& $1.6 \times 10^{-5}$\\

\hline
28 & Ours (STFT: magnitude, $L_1 \lambda=1$) & WS-Replicate &   0.75   &   --       & $2048 \times 1024$ & 0.85 & 26.15 & Full color & 31.24  &    $3.9 \times 10^{-4}$  \\
29 & \textbf{Ours (STFT: magnitude, $L_1 \lambda=1$)} & \textbf{WS-Replicate} &   0.75   & \checkmark  & $2048 \times 1024$ & \textbf{0.87} & 26.88 & Full color & \textbf{31.61} &    $3.9 \times 10^{-4}$  \\

30 & Ours (STFT: magnitude, DTW $\lambda=10^{-4}$) & WS-Replicate &   0.75   &   \checkmark      & $2048 \times 1024$ & \textbf{0.84} &  26.20 &  Full color &  30.46 & $5.1 \times 10^{-2}$\\

31 & Ours (STFT: magnitude, $L_1 \lambda=1$) & WS-Replicate &   0.5   &   \checkmark      & $2048 \times 1024$ &  0.82&  25.50 & Full color& 34.96& $2.6 \times 10^{-4}$\\

32 & Ours (STFT: magnitude, $L_1 \lambda=1$) & WS-Replicate &   0.9   &   \checkmark      & $2048 \times 1024$ &  \textbf{0.86}&   26.73&  Full color &27.90&$5.8 \times 10^{-4}$ \\

\hline
33 & Ours (STFT: magnitude, $L_1 \lambda=1$) & Multichannel &   0.75   &   --       & $2048 \times 1024$ & 0.83 & 23.26 & Partial color & 20.02  &    $2 \times 10^{-3}$ \\

\specialrule{.2em}{.1em}{.1em}
\end{tabular}}
\caption{\textbf{Results of the ablation study}. The metrics reported are SSIM and PSNR for image quality and SNR for audio quality. We also include qualitative information on color reconstruction (\emph{full color} represents the case when the whole spectrum of colors can be reconstructed, \emph{partial color} refers to the case when the color spectrum is reconstructed partially, \emph{no color} refers to only black and white image reconstruction from RGB images, and \emph{none} refers to no minimal image reconstruction, as hinted by the low SSIM values), and the waveform loss values (soft DTW or $L_1$, depending on the model). For reference, a signal with an SNR of 30 decibels (dB) or higher can be considered a perceptually clean signal \cite{ICSIspeechFAQ}. }
\label{table:all-results}
\end{table*}
\subsection{Higher stego resolution}\label{section:large-container}

Our baseline system assumes a \emph{stego} signal of size $1024\times512$, which is determined by the STFT applied on the input audio waveform with a set of hyperparameters (frame length and hop size). However, these values are arbitrary and could be changed, specifically, to increase the resolution of the \emph{stego} spectrogram. In this section we explore the possibilities offered by the use of a larger spectrogram, which should increase the embedding capacity of the \emph{stego} signal (and the robustness of the whole system if replication is used). Increasing the frame length of the STFT results in a larger size in the frequency dimension, while reducing the hop size of overlapping windows causes the container to increase in the time dimension. We applied both modifications to obtain a container of size $2048\times1024$, preserving the property of the dimensions being powers of two, allowing for efficient computations. Some adaptations have been required to accommodate for the larger container size:

\textbf{\emph{Stretch} method} needs to interpolate to a larger target size, the same as the \emph{stego} spectogram.

\textbf{\emph{Replicate-based} methods} use 8 replicas instead of 2, which are arranged in a $4 \times 2$ grid. As a consequence, \emph{W-Replicate} and \emph{WS-Replicate}, use 8 weights instead of 2 to scale each copy individually. \emph{WS-Replicate}'s decoder also needs to accept input of depth 8 instead of 2. 

\textbf{\emph{Multichannel}}'s encoder outputs 32 replicas instead of 8; the decoder also expects an input of depth 32. These are arranged in an $8 \times 4$ grid.
\begin{figure}[ht]
\centering
\includegraphics[width=\linewidth]{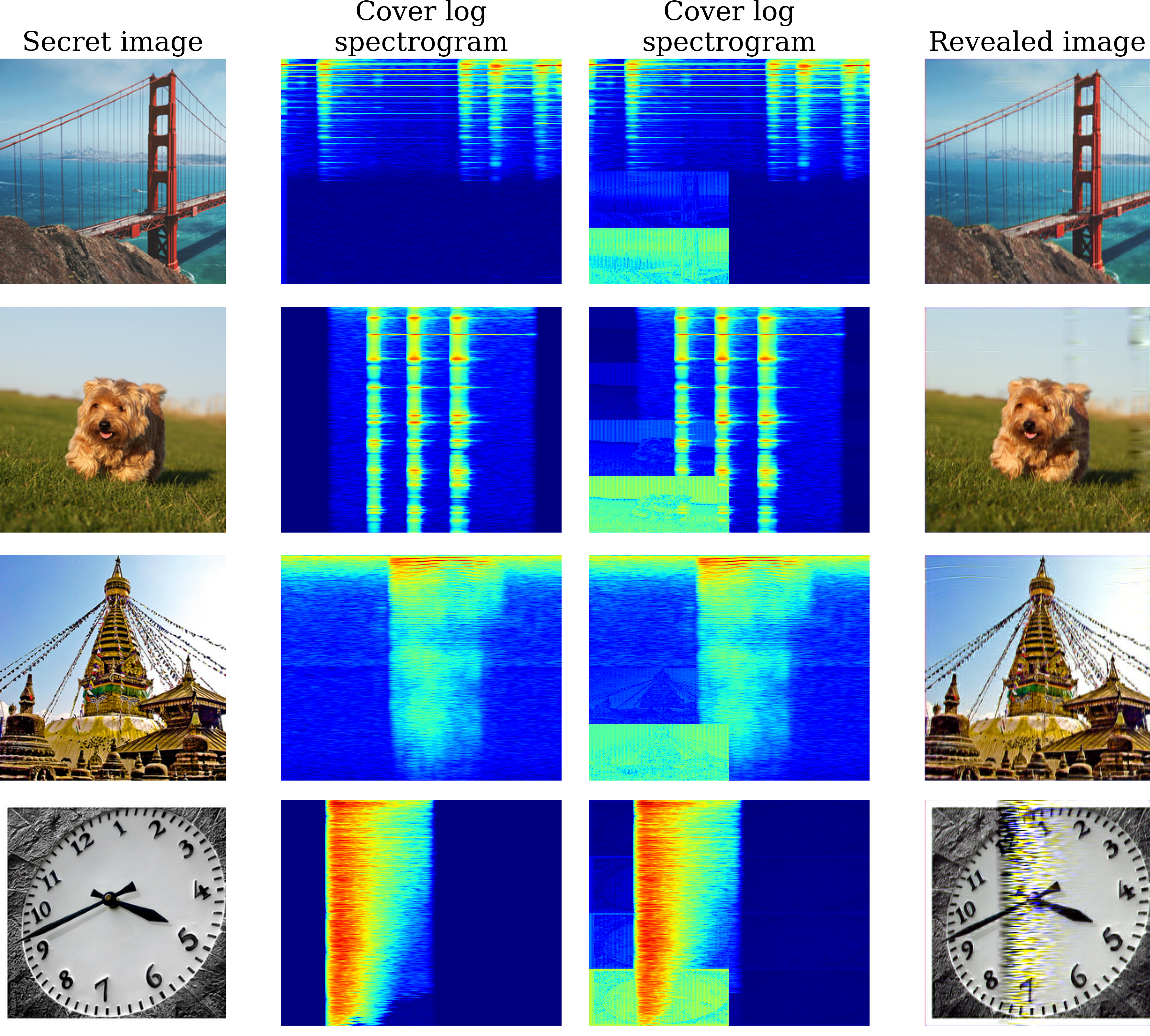} \caption{\textbf{Examples}. Several reconstruction samples with different images and audios using model \#29 from Table \ref{table:all-results}.} 
\label{fig:samples-pixinwav2}
\end{figure}

\subsection{Pixel-shuffle RGB channels with luma}

The pixel-shuffle operation \cite{pixelshuffle} used to flatten the image arranges each $1 \times 1 \times 3$ pixel into a $2 \times 2 \times 1$ grid. Thus, for every RGB pixel, we obtain a $2 \times 2$ grid where we can buffer the values. PixInWav \cite{geleta2021pixinwav} padded a value of 0 into the fourth component of every grid. Contrary to this zero-padding approach, we propose padding with the luma component of the pixel in question instead, as a way to add redundancy to the signal that can be later used for error correction in the decoder side. This peculiar pixel-shuffle step then outputs, for every pixel, a $2\times 2$ grid of 4 values $[R,G,B,Y]$, where $Y$ represents the luma component of the pixel, computed from the RGB components by a standardized transformation to the YCbCr color space \cite{jpeg-format}.
On the decoder side, for each $[R,G,B,Y]$ pixel, the YCbCr representation is computed from the RGB values. The newly computed $Y$ value is then averaged with the received $Y$, and the whole pixel is transformed back to the RGB color space to yield the final image.

\section{Experiments}\label{sec:cvpr2023-author_kit-v1_1-1/Experiments} 

\subsection{Datasets}

We have used a subset of 10,000 color images from the ImageNet Large Scale Visual Recognition Challenge 2012 (ILSVRC2012), sampling 10 images per ImageNet class, \cite{deng2009imagenet}. Every image has been cropped and scaled to $256 \times 256 \times 3$, normalized, and paired with the STFT transformation of an audio clip (roughly 1.5s) sampled at 44,100 Hz from the FSDnoisy18k dataset \cite{fsdnoisy18k}. The audio dataset contains a variety of different sounds, ranging across 20 different classes, among which we can find voice, music and noise. 

\subsection{Ablation Study}

In this section, we present the results of our ablation study to determine the impact of the proposed enhancements (Figure \ref{fig:arch}). Table \ref{table:all-results} summarizes the results of our experiments and Figure \ref{fig:samples-pixinwav2} displays a selection of visual examples. We provide a numerical identifier for each of the models for reference.

\textbf{STFT instead of STDCT}. We assessed the impact of using the magnitude of the STFT as a \emph{stego} signal instead of the STDCT spectrogram as in \cite{geleta2021pixinwav}. Comparing STDCT models \#1--\#5 against STFT models \#6--\#10 we can see that STDCT models struggle to find a balance point with good reconstruction of both, image and audio, while STFT models show a good performance in both.

\textbf{Modification of the loss function}. As a result of our choice of STFT over STDCT, we compared the results of computing the waveform loss using the $L_1$ distance instead of the soft DTW discrepancy. In Table \ref{table:all-results} we can compare DTW models \#6, \#18, \#22, \#25 and \#30, with $L_1$ models \#7, \#17, \#21, \#24 and \#29. Using the $L_1$ distance, results were slightly better in most cases. Note that the soft DTW loss proved superior when using the STDCT, as seen when comparing models \#4 and \#5, explained by the reasoning in Section \ref{sec:phase-loss-wav}.

\begin{figure*}[ht]
\centering
\includegraphics[width=\linewidth]{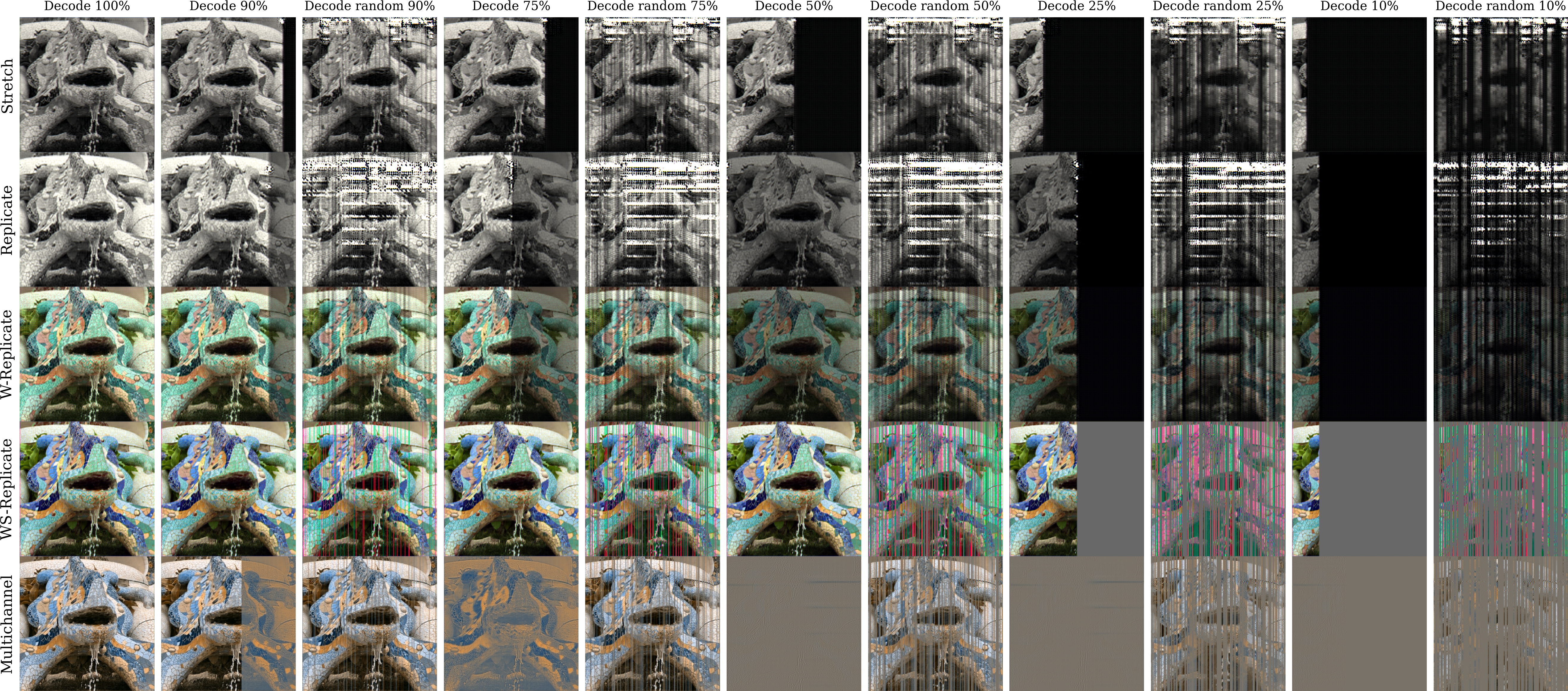}
\caption{\textbf{Comparison of different embedding methods with respect to robustness}. We have tried to decode smaller temporal segments containing image information to see how much distortion is induced if some part of the \emph{stego} signal is lost. We explored decoding different sequential and random percentages of the spectrogram on the temporal axis.}
\label{fig:robustness-experiments}
\end{figure*}

\textbf{Type of STFT \emph{stego} signal}. Next, we compared the performance of the steganographic operation based on the kind of the \emph{stego} signal: just magnitude, just phase, or a combining both. We find that using the phase as the sole container is clearly inferior to using the magnitude (compare models \#9 and \#10 in Table \ref{table:all-results}), as there is a very significant drop in both image and audio reconstruction quality. Our reasoning for these results is twofold. Firstly, the phase is a much noisier signal in nature, which makes the task of hiding information more difficult. Secondly, minor modifications to the phase component result in a more perceptible distortion in the reconstructed audio, thus rendering the task of concealing the \emph{secret} signal more challenging. Finally, we compared using both the magnitude and phase as \emph{stego} signals simultaneously (model \#11). The results from Table \ref{table:all-results} show that while using both \emph{stego} signals does substantially increase image quality, there is a significant drop in audio quality, possibly as a consequence of additionally distorting the phase. 
In conclusion, our study suggests that it is not worth using the phase as a \emph{stego} signal, since it does not improve the metrics obtained with the baseline model that only uses the magnitude. Therefore, there is no justification for the added overhead in the model. 

\textbf{Comparison of embedding methods}. A comparison of different embedding methods has been conducted, and the results are presented in Table \ref{table:all-results} accross models \#11--\#15 (\emph{stego} signal of size $1024 \times 512$) and models \#16, \#20, \#23, \#28, \#33 (\emph{stego} signal of size $2048 \times 1024$). The \textit{Multichannel} method exhibits a considerable enhancement in image quality, albeit at the cost of a significant decrease in audio metrics, thereby rendering it less practical for most real-world applications. Conversely, all replicate-based embedding methods outperform the baseline \textit{Stretch} approach. Qualitative assessments depicted in Figure \ref{fig:embedding-methods} demonstrate that \textit{WS-Replicate} can generate a superior reconstruction of the original image, as it is the only method capable of preserving the authentic color. 

\textbf{Buffering the luma component in the pixel shuffle operation}. Experiments shown by pairs of models \#16 and \#17, \#20 and \#21, \#23 and \#24, \#28 and \#29 show that this addition does improve the quality of the revealed images while maintaining a comparable audio quality with regard to the baseline model.

\textbf{Higher resolution \emph{stego} signal}. The values from Table \ref{table:all-results} show a very significant improvement when using a larger resolution of the \emph{stego} signal (compare models \#6--\#15 against \#16--\#33), both in image and audio quality, as it is expected from having more capacity for carrying information. This increase in performance comes at the cost of increased memory usage and longer training times. Note that, for training purposes, the audio transform can be precomputed for every audio; however, the inverse transform is still needed to compute the waveform loss.

\subsection{Embedding method effect on robustness}
To evaluate the effect of different embedding methods on the robustness of the steganographic method, we attempted to decode smaller temporal segments of the \emph{stego} signal. In our experiment, we selectively zero out the spectral content at different time frames of the \emph{stego} spectrograms, simulating a scenario where some data is lost during transmission \cite{e24070878, Gopalan2004STEGANOGRAPHYFC}, either in large contiguous chunks or at random positions.

The qualitative results can be appreciated in Figure \ref{fig:robustness-experiments}. They show that methods that use replication are more robust than the baseline \emph{Stretch} approach, allowing to recover most of the image even if a large part of the \emph{stego} signal is lost.

\subsection{Computational cost study}

\begin{table}[]
\centering
\resizebox{0.75\linewidth}{!}{\begin{tabular}{l|ll}
\specialrule{.15em}{.075em}{.075em} 
Enhancement                    & \# params         & GMAC $\downarrow$                  \\ 
\specialrule{.15em}{.075em}{.075em} 
Baseline                       & 962128            & 34.6                  \\ \hline
STFT (magnitude)                     & +0                & +0\%                  \\
STFT (phase)                   & +0                & +0\%                  \\
STFT (magnitude + phase)               & +962131           & +100.00\%             \\
$L_1$ loss                        & +0                & +0\%                  \\
Replicate                      & +0                & +0\%                  \\
W-Replicate                    & +4                & +0\%                  \\
WS-Replicate                   & +584              & -32.89\%              \\
Multichannel                   & +12735            & -81.68\%              \\
Stretch (large)                & +0                & +200.00\%             \\
Replicate (large)              & +0                & +200.00\%             \\
W-Replicate (large)            & +4                & +200.00\%             \\
WS-Replicate (large)           & +4103             & -30.23\%              \\
Multichannel (large)           & +67695            & -73.20\%              \\
Luma                           & +0                & +0\%                  \\ 
\specialrule{.15em}{.075em}{.075em} 
\end{tabular}}
\caption{\textbf{Breakdown of computational costs}. In this table, we show the relative increment in the number of parameters and Giga-Multiply–accumulate operations (GMAC) of each of the proposed enhancements with respect to the baseline model.}
\label{table:comp-load}
\end{table}

The results from the previous sections show that some setups obtain better performance than others. However, in some cases this comes at the cost of increased computational load, during both training and inference. The trade-off between these two factors should take into account the available resources and future use of the model. Table \ref{table:comp-load} presents our results of such analysis. An increase in the number of parameters implies a higher memory usage and longer execution times, and an increase in Giga Multiply–accumulate operations (GMAC) operations generally indicates longer execution times and energy consumption. Among models with similar performance, lower values in both metrics should be preferred.

\textbf{Cost of using the STFT instead of STDCT}. For both transforms there exist efficient algorithms with equal asymptotic complexity \cite{Cooley_Tukey_1965, dct_complexity_1977}. We thus consider the two options to be equal in this regard. However, there is an additional cost if we use the magnitude and phase together as \emph{stego} sigals -- this scenario doubles the number of parameters and MAC operations, since two separate encoder and decoder networks are used (plus a small coupling network). 

\textbf{Choice over $\bm{L_1}$ and DTW}. The usage of the $L_1$ loss instead of dynamic time warping cannot be directly assessed, since this is only used during the training process and outside the model. It should be noted, however, that the $L_1$ loss is generally much more efficient to compute ($\mathcal{O}(n)$ time) than (soft-)DTW ($\mathcal{O}(n^2)$ time) \cite{softdtw}.

\textbf{Cost of embedding methods}. The different embedding methods can also be compared in terms of computational load (Table \ref{table:comp-load}). \emph{Replicate} does not add any additional parameters with respect to the baseline \emph{Stretch} method, the only difference being that the information is duplicated and concatenated instead of being upsampled. \emph{W-Replicate} adds four additional parameters that scale each of the two replicas (which is done at both, the encoder and decoder end). The effect on the load is negligible. \emph{WS-Replicate} and \emph{Multichannel} methods use deeper convolution kernels over smaller resolution tensors, thus increasing the amount of parameters while decreasing the total number of operations. This is especially noticeable in \emph{Multichannel}.

\textbf{Using higher resolution of \emph{stego} signal}. When using a larger \emph{stego} spectrogram, the amount of parameters remains the same, except for \emph{WS-Replicate} and \emph{Multichannel}, that use even deeper kernels to process a larger number of replicas. However, the number of floating-point operations triples.

\textbf{Cost of buffering the luma component}. The usage of the luma channel in the pixel-shuffle operation only entails a color space change (done through a single matrix multiplication) and averaging the two luma values. These operations do not add any extra trainable parameters, and the computational cost is negligible.

\vspace{-1mm}
\section{Discussion and Conclusion}

We have presented a set of key enhancements for an existing image-in-audio deep steganography method, among which: the use of STFT, introduction of redundancy in the encoding and decoding steps for error correction, and buffering of additional information in the pixel subconvolution operation. Our experiments have demonstrated that our approach outperforms the existing method in terms of robustness and  perceptual transparency. Our novel approach, thus, represents a significant step forward in the field of multimodal deep steganography, promising improved security and confidentiality in a wide range of applications. 

Our qualitative results show a clear system bias to distort those parts of the image where the \emph{cover} spectrogram exhibits high magnitudes. Although redundancy through replication ameliorates this issue partially, primarily by concealing a substantial portion of the information in the higher frequencies, where the degree of distortion is typically lower, it proves inadequate in certain scenarios where the \emph{cover} spectrogram manifests high values across all frequencies for a brief duration.

Future work can explore new techniques to increase the system's robustness on these rare cases, and also the applicability of our approach in real world scenarios, exposing the \emph{stego} signal to acoustic alterations, such as ambient noise and reverberations.  

\subsubsection*{Acknowledgements}
We express our gratitude to Pau Bernat Rodríguez for his discussions throughout this study and his contributions to the project codebase.

{\small
\bibliographystyle{ieee_fullname}

}

\end{document}